\documentclass[a4paper,portrait]{iopart}
\usepackage{epsfig}
\usepackage{graphicx}
\begin{document}
\title{Photodisintegration studies on p-nuclei: The case of Mo and Sm isotopes}
\author{
C Nair\dag, A R Junghans\dag, M Erhard\dag, D Bemmerer\dag, R
Beyer\dag, P Crespo\dag, E Grosse\dag\ddag, M Fauth\dag, K
Kosev\dag, G Rusev\dag, K D Schilling\dag, R Schwengner\dag, and A
Wagner\dag}
\address{\dag
\ Institut f\"ur Strahlenphysik,
    Forschungszentrum Dresden-Rossendorf,
    Postfach 510119, 01314 Dresden, Germany}
\address{\ddag\ Institut f\"ur Kern- und Teilchenphysik,
    Technische Universit\"at Dresden,
    01062 Dresden, Germany}
\ead{chithra.nair@fzd.de}
\begin{abstract}
In explosive stellar environments like supernovae, the temperatures
are high enough for the production of heavy neutron-deficient
nuclei, the so-called p-nuclei. Up to now, the knowledge of the
reaction rates of p-nuclei is based on theoretical parameterizations
using statistical model calculations. At the bremsstrahlung facility
of the superconducting electron accelerator ELBE of FZ
Dresden-Rossendorf, we aim to measure the photodisintegration rates
of heavy nuclei experimentally. Photoactivation measurements on the
astrophysically relevant p-nuclei $^{92}$Mo and $^{144}$Sm have been
performed with bremsstrahlung end-point energies from 10.0 to 16.5
MeV. First experiments on the short-lived decays following the
reaction $^{144}$Sm($\gamma$,n) are carried out using a pneumatic
delivery system for rapid transport of activated samples. The
activation yields are compared with calculations using cross
sections from recent Hauser-Feshbach models.
\end{abstract}
\submitto {JPG}
\section{Introduction}\label{introduction}
The nuclei heavier than iron (Z $>$ 26) are synthesized mainly by
neutron-capture reactions - the astrophysical r- and s- processes.
The 35 neutron deficient stable isotopes between Se and Hg that are
shielded from the rapid neutron capture by stable isobars are
classically referred to as the p-nuclei. From the current
understanding, best possible sites for the production of  p-nuclei
are O-Ne rich layers of Type II supernova explosions. Their
production mechanism is understood as chains of photodisintegrations
like ($\gamma$,n), \mbox{($\gamma$,p)} and ($\gamma$,$\alpha$) on r-
or s- seed nuclei~\cite{Woosley1978,Rayet1990,Lambert1992}. The
solar system isotopic abundances of the p-nuclei are generally only
in the order \mbox{(0.01-1)\%}. For an overview on the experimental
techniques, status and problems associated with p-nuclei studies,
see~\cite{Arnould2003, Mohr2007}.

For p-process nucleosynthesis modelling, precise knowledge of the
astrophysical reaction rates is important. Despite the efforts in
recent years, very few experimental data for photodisintegration cross
sections exist. Therefore, the reaction rates presently used are based
on the cross sections obtained from Hauser-Feshbach
statistical model calculations. The aim of our experiments
with real photons is to determine the cross sections experimentally
to test the statistical model calculations.

The present paper focuses on the results from photodisintegration
studies of the p-nuclei $^{92}$Mo and $^{144}$Sm. The nuclide
$^{144}$Sm has been in the frame of the p-process chronometer
$^{146}$Sm~\cite{Audouze1972}. There were several efforts to
determine the $^{146}$Sm/$^{144}$Sm production ratio experimentally
which varies due to uncertainties in different inputs entering into
the calculation, see~\cite{Somorjai1998}. One of the important
nuclear physics input is the proper optical potential at energies of
astrophysical relevance which has been derived from
$^{144}$Sm($\alpha$,$\alpha$)$^{144}$Sm elastic
scattering~\cite{Mohr1997}. We have investigated the
$^{144}$Sm($\gamma$,$\alpha$)$^{140}$Nd reaction via photoactivation
method  for the first time. With the rapid transport system for
activated samples, the short-lived nuclides following the reaction
$^{144}$Sm($\gamma$,n)$^{143}$Sm were identified.
Photodisintegration studies on $^{92}$Mo are performed aiming to
test the underprediction of the Mo species in the network
calculations. Preliminary results from $^{92}$Mo data have been
presented~\cite{ErhardEPJ,Erhard2006}.
\section{Experimental Setup}\label{setup}
In order to study the photodisintegration reactions on $^{92}$Mo and
$^{144}$Sm, photoactivation measurements are being performed at the
bremsstrahlung facility of the superconducting electron accelerator
ELBE (Electron Linear accelerator of high Brilliance and low
Emittance) of FZ Dresden-Rossendorf~\cite{Schwengner2005,
Wagner2005}. ELBE delivers electron beams of energies up to 20 MeV
with average currents up to 1 mA which is appropriate for probing
photon-induced reactions.

The primary electron beam is focussed onto one of the several
niobium radiators with thicknesses varying between 1.7~mg/cm$^2$ and
10~mg/cm$^2$ corresponding to $1.6\cdot10^{-4}$ and $1\cdot10^{-3}$
radiation lengths. Behind the radiator, the electron beam is
separated from the photons by a dipole magnet and dumped into a
graphite cylinder with a conical recess. The length of the cylinder
is 600 mm and diameter is 200 mm. The photoactivation site is
located behind the beam dump where available photon fluxes amount up
to  10$^{9}$ cm$^{-2}$ s$^{-1}$ MeV$^{-1}$, see Station B in Fig. 1.

In this high-flux photoactivation area, targets of Mo or Sm were
irradiated together with Au samples. At the same time another Au
sample was irradiated together with a $^{11}$B sample at the
photon-scattering site, which is separated from the high-flux site
by a heavy-concrete wall. A 2.6 m long collimator placed in the wall
shapes the photon beam that hits the photon-scattering target with a
photon flux of about 10$^{8}$ cm$^{-2}$ s$^{-1}$ MeV$^{-1}$. Photons
scattered from $^{11}$B were measured with four high-purity
germanium (HPGe) detectors surrounded by escape-suppression shields
consisting of bismuth-germanate (BGO) scintillation detectors. The
photon flux at the photon-scattering site was determined by means of
the known integrated cross sections of the states in $^{11}$B
depopulating via $\gamma$ rays. Using the known cross section for
the $^{197}$Au($\gamma,n$) reaction and the activities of the Au
samples at the photon-scattering site and the high-flux site, the
activation yields of the Mo and Sm isotopes were deduced.

The molybdenum targets used were of natural abundance (discs with
mass 2-4 g and diameter 20 mm). For the activation of samarium we
used fine Sm$_2$O$_3$ powder filled in polyethylene capsules (mass
3-4 g and diameter 18 mm).  Au samples were of mass $\approx$ 200 mg
and were irradiated both in the photoactivation site and  in
brems\-strahlung cave.

The number of activated nuclei produced during the activation was
determined offline by measuring the decay in a low-level counting
setup by HPGes with an efficiency 90\% or 60\% relative to a
3"$\!\times$3" NaI detector.
\section{The Rabbit System}\label{rabbitsystem}
\begin{figure}
\begin{center}
\resizebox{\textwidth}{!}{\epsfig{file=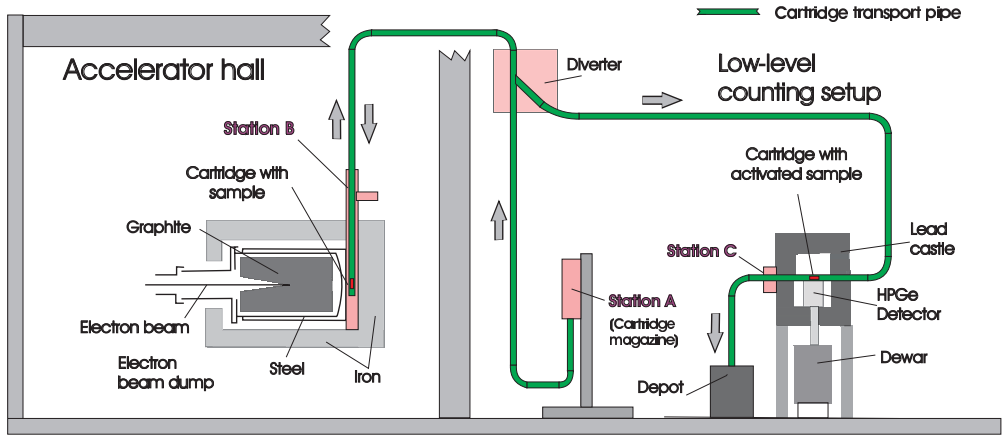,width=\textwidth,angle=270}}
\end{center}
\caption{\label{fig:rabbitsys}Pneumatic delivery system for fast
transport of activated samples from irradiation site to the decay
measurement site. The different stations for sample loading (A),
irradiation (B) and decay measurement (C) is shown.}
\end{figure}
An air-driven pneumatic delivery system (rabbit system) has been
built for experiments with short-lived isotopes resulting from
activation. A sketch of the setup is shown in Fig.
\ref{fig:rabbitsys}. The whole system uses compressed air to
transport the samples through polyamide (PA) tubes with a diverter
making way to the to-and-fro movement of the sample cartridges. The
samples to be irradiated are enclosed in polyethylene cassettes and
loaded in Station A. They are shot to the high photon flux area
behind the vacuum steel vessel aligned on the axis of the electron
beam (Station B). After irradiation, the samples are transported
within about 15 s to the lead-shielded low-level counting setup
where the decay is measured with a coaxial HPGe detector. The loss
time of 15 seconds arises from the time taken for transport of the
sample plus placing the sample center above the HPGe detector. The
variation in horizontal positioning of the samples by the rabbit
system is very small, as the  samples are stopped and then pushed
back into position by a slowly operating pneumatic feedthrough.
After the decay measurement, samples are shot to a radiation
shielded container (depot, see figure).

\begin{figure}
\begin{center}
\epsfig{file=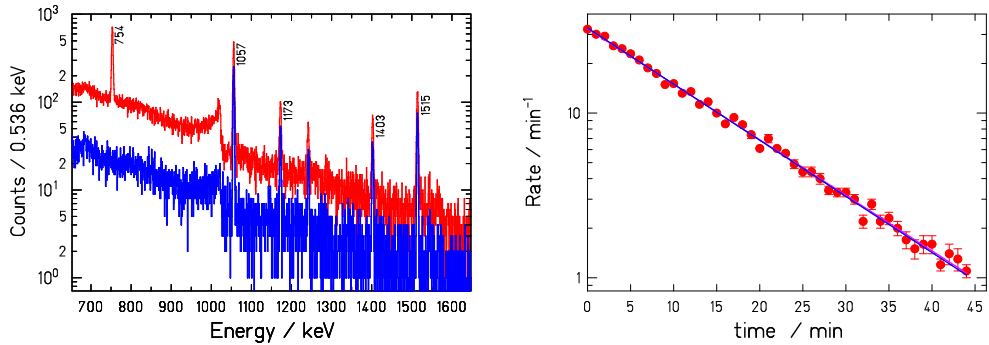,angle=270,width=\textwidth}
\end{center}
\caption{\label{fig:spectra}Decay spectra following the reaction
$^{144}$Sm($\gamma$,n). The $\gamma$-ray energies are marked for the
transitions from the decay of the isomeric and ground states of
$^{143}$Sm. On the right panel, exponential decrease in the rate of
$^{143}$Sm identified with the strongest line at 1057 keV is shown
which was used for the half-life determination.}
\end{figure}

The gamma-ray spectra for an activated sample of Samarium after 10
minutes of irradiation using the rabbit system is shown in Fig.
\ref{fig:spectra}. In the overlayed figure on the left panel, upper
spectrum was taken 1 minute after an irradiation and the lower one
after 9 minutes. The decays following
$^{144}$Sm($\gamma$,n)$^{143m}$Sm and
$^{144}$Sm($\gamma$,n)$^{143}$Sm reactions with half-lives 66 s and
8.75 min respectively are clearly seen. The
$^{144}$Sm($\gamma$,n)$^{143m}$Sm reaction is identified with the
unique line at 754 keV and the $^{144}$Sm($\gamma$,n)$^{143}$Sm with
the lines above 1000 keV.  On the right panel, exponential decay of
the ground state of $^{143}$Sm is shown. The measured half-life is
8.83$\pm$ 0.08 min which is in good agreement with the recent
literature value 8.75$\pm$0.08 min~\cite{nndchalflife}.
\section{Data Analysis}\label{analysis}
In activation experiments, the sample to be studied (target) is
irradiated always with a gold sample as normalization standard. The
reaction $^{197}$Au($\gamma$,n) is well suited as a standard for
activation experiments and has been measured previously by different
methods~\cite{Vogt2002,Veyssiere1970,Berman1987}. The number of
$^{196}$Au nuclei produced during the activation is determined by
counting the gamma decay in  the shielded detector setup. The
absolute efficiency of the HPGe detectors is based on measurements
with several calibration sources from PTB and Amersham (systematic
uncertainty in activity 0.6-1.5\%) in the energy range from 0.12 to
1.9 MeV~\cite{PTB}. The sources were positioned at the same distance
as the center of the irradiated samples. The effect of source
extension and self-absorption of gamma-rays was determined with
EFFTRAN~\cite{Vidmar2005} and GEANT3~\cite{GEANT3} simulations. For
the counting geometry directly on top of the detector the efficiency
of the Sm$_2$O$_3$ samples is reduced by 3\% compared to the point
source value, whereas for the rabbit-system counting geometry the
efficiency is reduced by 2\%. Coincidence summing effects are
minimized by using a Cd absorber with 1.5 mm thickness. They depend
strongly on the decay scheme. The three dominant decay transitions
of $^{196}$Au used for analysis are at 333, 356 and 426 keV. For the
transition at 333keV, the coincidence summing correction amounts to
24\% and for 356 keV it is 6\% both with a relative uncertainty of
5\%.

The number of radioactive nuclei $N_{\mathrm{act}}(E_0)$
produced in a photo-activation experiment is proportional to the
integrated product of the absolute photon flux
$\Phi_{\gamma}(E,E_0)$ and the photodisintegration cross section
$\sigma _{\gamma ,\mathrm{x}}(E)$ with the integral limits from the reaction
threshold energy $E_{\mathrm{thr}}$ up to the bremsstrahlung spectrum
end-point energy $E_0$. The symbol $\mathrm{x = n, p, \alpha }$ denotes the
emitted particle.
\begin{equation}
N_{\mathrm{act}}(E_0) =  N_{\mathrm{tar}} \cdot
\int_{E_{\mathrm{thr}}}^{E_{\mathrm{0}}} \sigma_{\mathrm{\gamma
,x}}(E)\cdot \Phi_{\gamma}(E,E_0)\,dE \label{eq:yi}
\end{equation}
After irradiation, the number of radioactive nuclei $N_{act}(E_0)$
is determined experimentally by
measuring the decay using a HPGe detector and the formula reads:
\begin{equation}
N_{\mathrm{act}}(E_0) = \varepsilon^{-1}(E_{\gamma}) \cdot
N_{\gamma}(E_{\gamma},E_0) \cdot p^{-1}(E_\gamma) \cdot
\kappa_{\mathrm{corr}} \label{eq:yint}
\end{equation}
$ N_{\gamma}(E_{\gamma},E_0),\varepsilon(E_{\gamma}), p(E_\gamma)$
stand for the dead-time and pile-up corrected full-energy peak
counts of the observed transition, the absolute efficiency of the
detector at the energy $E_{\gamma}$ and the emission probability of
the photon with energy $E_\gamma$ respectively. The factor $\kappa_{\mathrm{corr}}$
accounts for the decay losses during irradiation, in between
irradiation and decay measurements and during the measurement.

The activation yield is denoted by  $Y_{\mathrm{act}}$ and is expressed as
the ratio of the number of activated nuclei to the number of target
atoms in the sample. For example, for the $^{92}$Mo ($\gamma$,$\alpha$) reaction,
\begin{equation}
 Y_{\mathrm{act}} = N_{\mathrm{act}}(^{92}\rm{Mo} (\gamma
,\alpha))/ N_{\mathrm{tar}}(^{92}\rm{Mo})
 \end{equation}
The experimental data are compared to the yield integrals calculated
with a simulated thick-target bremsstrahlung spectrum and
photodisintegration cross sections predicted by Hauser-Feshbach
models~\cite{Rauscher2004,Koning2005}.
\section{Results and Discussion}\label{results}
\begin{figure}
\begin{center}
\epsfig{file=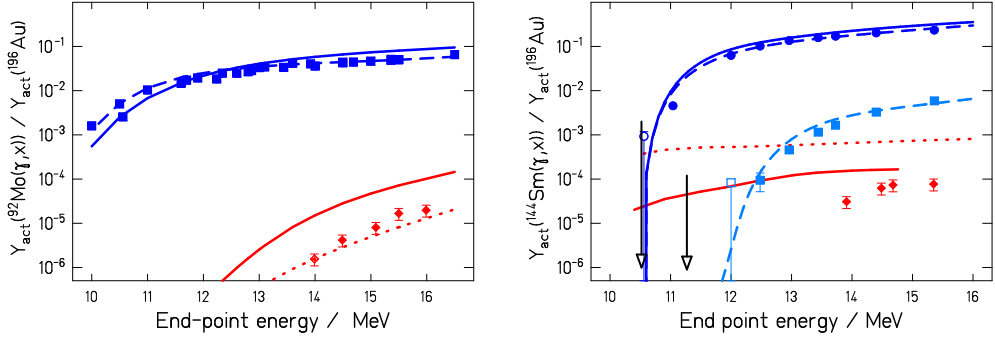 ,angle=270,width=\textwidth}
\end{center}
\caption{\label{fig:actyield} Experimental activation yields for
photodisintegration reactions in $^{92}$Mo (left) and $^{144}$Sm
(right) isotopes normalized to the  $^{197}$Au($\gamma$,n) yield.
For $^{92}$Mo, the $^{92}$Mo(($\gamma$,p)+($\gamma$,n)) (squares)
and $^{92}$Mo($\gamma$,$\alpha$) (diamonds) yields are shown. On the
right panel, the $^{144}$Sm($\gamma$,n) yields from  $^{143}$Sm
(circles) and $^{143m}$Sm (squares) decay are plotted together with
the $^{144}$Sm($\gamma$,$\alpha$) yield (diamonds). The effective
neutron separation energies for the respective
$^{144}$Sm($\gamma$,n) reactions are marked by arrows and open
symbols denote the detection threshold for the lowest measured
energies. The dashed and dotted lines denote yield calculations
using cross sections from TALYS~\cite{Koning2005} for the respective
($\gamma$,n) and ($\gamma$,$\alpha$) reactions. The solid lines
stand for predictions with cross sections from
NON-SMOKER~\cite{Rauscher2004} code for both reactions.}
\end{figure}
Measured activation yields relative to the $^{197}$Au($\gamma$,n)
reaction yield for photodisintegration reactions in $^{92}$Mo and
$^{144}$Sm are shown in Fig. \ref{fig:actyield}. In  $^{144}$Sm, the
neutron separation energy is 10.5 MeV and the proton separation
energy 6.7 MeV. The $^{144}$Sm($\gamma$,n) reaction produces
$^{143}$Sm or $^{143m}$Sm. Since both of them are short-lived the
irradiation was carried out using the rabbit system, see Sect.
\ref{rabbitsystem}. The measured activation yield for
$^{144}$Sm($\gamma$,n)$^{143}$Sm and
$^{144}$Sm($\gamma$,n)$^{143m}$Sm reactions relative to the standard
$^{197}$Au($\gamma$,n) yield agrees within 20\% to the simulated
yield integrals with cross sections predicted by theoretical
models~\cite{Rauscher2004,Koning2005}.

The $^{144}$Sm($\gamma$,$\alpha$) reaction was identified by the
transition at 1596 keV from $^{140}$Pr which is the short-lived
daughter of the $(\gamma,\alpha)$ reaction product $^{140}$Nd. For
$^{144}$Sm the $Q(\alpha)$ value is -0.145 MeV. Half-lives for
$^{140}$Nd and $^{140}$Pr are 3.37 days and 3.4 minutes
respectively. Measured reaction yields compared to the calculated
ones are shown in Fig. \ref{fig:actyield}. The preliminary
experimental data for $^{144}$Sm($\gamma$,$\alpha$) relative yield
is much below the simulated values.The theoretical
models~\cite{Rauscher2004,Koning2005} in this case differ strongly
which could be an indication for the different $\alpha$-nucleus
potentials entering into the calculation. The predictions given by
the two models are also dependent on the different parametrization
of  dipole strength functions, level densities  and the mass models
used. The experiment on $^{144}$Sm($\gamma$,p) reaction is in
progress.

In $^{92}$Mo, the experimental yields for
$^{92}$Mo(($\gamma$,p)+($\gamma$,n)) agree with the simulations
using predicted cross sections from~\cite{Rauscher2004,Koning2005}
within 20\%. For the $(\gamma,\alpha)$, predictions from
~\cite{Koning2005} agree within a factor of 2 whereas the one from
~\cite{Rauscher2004} is much above the experimental value.

\ack We thank the ELBE-team for providing the stable beam, J.
Claussner and his co-workers for building the rabbit system and A.
Hartmann for continuing technical assistance.

\end{document}